\documentclass[12pt,aps,tightenlines,notitlepage,showpacs,superscriptaddress]{revtex4-1}
\usepackage[dvips]{epsfig}

\usepackage{amsmath}
\usepackage{amssymb}
\usepackage{bm}
\usepackage[mathscr]{eucal}
\usepackage{theorem}
\usepackage{graphicx}
\usepackage{psfrag}
\usepackage{color}
\usepackage{wasysym}
\include{graphix}
\usepackage{framed}
\usepackage{enumitem}%
\usepackage{lipsum}
\usepackage{float}
\usepackage{MnSymbol}

\usepackage{caption,subcaption}
\usepackage{mwe}
\usepackage{url}

\usepackage{tikz}
\usetikzlibrary{shapes}
\usepackage{xspace}

\graphicspath{{figures/}}

\newcommand{\vecx}{\mathbf{x}}
\newcommand{\vecy}{\mathbf{y}}
\newcommand{\vecz}{\mathbf{z}}

\newcommand{\vecb}{\mathbf{b}}
\newcommand{\vecrhs}{\mathbf{f}}
\newcommand{\vecrhsT}{\mathbf{d}}

\newcommand{\vecf}{\mathbf{f}}
\newcommand{\vecg}{\mathbf{g}}

\newcommand{\mxA}{\mathbf{A}}

\newcommand{\mxL}{\mathbf{L}}
\newcommand{\mxR}{\mathbf{R}}
\newcommand{\mxU}{\mathbf{u}}
\newcommand{\mxV}{\mathbf{v}}

\newcommand{\myqed}{\nobreak \ifvmode \relax \else
      \ifdim\lastskip<1.5em \hskip-\lastskip
      \hskip1.5em plus0em minus0.5em \fi \nobreak
      \vrule height0.75em width0.5em depth0.25em\fi}

\begin{document}
\title{Efficient Interleaved Batch Matrix Solvers for CUDA}
\author{Andrew Gloster}
\affiliation{School of Mathematics and Statistics, University  College Dublin, Belfield, Dublin 4, Ireland}
\author{Enda Carroll}
\affiliation{School of Mathematics and Statistics, University  College Dublin, Belfield, Dublin 4, Ireland}
\author{Miguel Bustamante}
\affiliation{School of Mathematics and Statistics, University  College Dublin, Belfield, Dublin 4, Ireland}
\author{Lennon \'O N\'araigh}
\affiliation{School of Mathematics and Statistics, University  College Dublin, Belfield, Dublin 4, Ireland}
\date{\today}

\date{\today}

\begin{abstract}
In this paper we present a new methodology for data accesses when solving batches of Tridiagonal and Pentadiagonal matrices that all share the same LHS matrix.
By only storing one copy of this matrix there is a significant reduction in storage overheads and the authors show that there is also a performance increase in terms of compute time.
These two results combined lead to an overall more efficient implementation over the current state of the art algorithms cuThomasBatch and cuPentBatch, allowing for a greater number of systems to be solved on a single GPU.

\end{abstract}

\maketitle



\vspace{0.1in}
\noindent{\bf{Program Summary}} \\
Program Title: CUDA Batched Tridiagonal and Pentadiagonal Schemes \\
Licensing Provision: Apache License 2.0 \\
Programming Languages: C, C++, CUDA \\
Computer: Variable, equipped with CUDA capable GPU \\
Operating System: Linux, Mac and Windows \\
Nature of Problem: Various implementations of batched Pentadiagonal and Tridiagonal solvers exist for CUDA using an interleaved data layout format.
All of the current state of the art implementations require large amounts of RAM due to the need for every thread to have its own copy of the LHS matrix.
There are many situations, particularly in the batch solving of PDEs in 1D and 2D, where the LHS matrix is the same, thus the data overheads are unnecessarily large reducing the number of systems that could be solved on a single GPU.\\
Solution method: In this paper we eliminate the LHS matrix requirements and allow every thread to access the LHS matrix, a dramatic saving in memory and also grants an additional speed--up over the existing state of the art.  \\
Tridiagonal Functions Source: \url{https://github.com/EndCar808/cuThomasConstantBatch} \\
Pentadiagonal Functions Source: \url{https://github.com/munstermonster/cuPentConstantBatch} \\

\section{Introduction}
\label{sec:intro}
Batched solutions of $\mxA \vecx = \vecb$, in particular where $\mxA$ is a tridiagonal or pentadiagonal matrix, are becoming increasingly prevalent methods for tackling a variety of problems on GPUs which offer a high level of parallelism~\cite{zhang2010fast, kim2011scalable, cuThomasBatch, cuPentBatch}. 
In the field of gravitational wave data analysis, much work is done simplifying complex waveforms and analyses to achieve results in a physically reasonable and desirable amount of time. A large portion of this work involves using cubic splines for interpolation~\cite{splineIEEE}, a highly parallelizable process that has shown promising results using GPUs in terms of accelerating established analysis procedures, as well as allowing for new methods with the extreme increase in computational speed.  
Fluid mechanics has also seen a broad application of GPUs where solutions of Poisson's equation are commonly required~\cite{hockney1964fast, valeroPoisson}, batch solving of 1D partial differential equations~\cite{cuPentBatch} and ADI methods for 2D simulations \cite{ADIGPU, gloster2019custen}.
Other examples of areas using GPUs include image in-painting \cite{imageInpainting}, simulations of the human brain~\cite{brainPaper} and animation~\cite{pixar}.
In this paper we will focus on the development of tridiagonal and pentadiagonal batch solvers with single LHS, multiple RHS, matrices in CUDA for application in solving batches of 1D problems and 2D ADI methods.
CUDA, an API extension to the C/C++ language, allows programmers to write code to be executed on NVIDIA GPUs.
In particular we develop solvers in this work which are more efficient in terms of data storage than the state of the art, this saving of data usage is due to the fact we have only one global copy of the LHS matrix rather than one for each thread/system.
While the primary improvement is in data storage reduction the new functions also provide increased speedup, due to better memory access patterns, when compared to the existing state of the art.
This increase in efficiency is also beneficial as it leads to further savings in resources, both in terms of run--time and electricity usage.
The latter of which is of increasing concern as modern supercomputers become increasingly bigger, requiring more and more electricity which has both financial and environmental implications.
The need for both energy efficient hardware and algorithms is becoming an ever more prevalent trend.

To date all existing batch pentadiagonal solvers in CUDA require that each system in the batch being solved have its own copy of the LHS matrix entries, leading to a sub-optimal use of GPU RAM and increased memory access overhead when only one global copy is actually needed. 
We also note that, while there are options for single LHS, multiple RHS, tridiagonal matrices in cuSPARSE these all rely on Cyclic Reduction or pivoting algorithms.
In the context of numerically solving PDEs on well behaved uniform grids these solution methods have unnecessary computational overhead when compared to the standard Thomas Algorithm for solving tridiagonal systems.
Thus we propose the implementation of two solvers, one implementing the Thomas Algorithm for tridiagonal matrices and the other implementing the pentadiagonal equivalent as presented in~\cite{cuPentBatch, numalgC} , each with a single global copy of the LHS matrix and multiple RHS matrices.
We will benchmark these implementations against cuThomasBatch \cite{cuThomasBatch} (implemented as gtsvInterleavedBatch in the cuSPARSE library) and existing work by the authors cuPentBatch \cite{cuPentBatch}, for tridiagonal and pentadiagonal matrices respectively, which are the existing state of the art for the algorithms we are interested in.
We point the user to the papers \cite{cuThomasBatch} and \cite{cuPentBatch} for existing benchmark comparisons of these algorithms with multiple LHS matrices with the cuSPARSE library and comparisons with serial/OpenMP implementations.
Thus we can compare our new implementations to the existing state of the art, cuThomasBatch and cuPentBatch, from which relative performance compared to the rest of cuSPARSE can be interpolated by the reader. 

The paper is laid out as follows, in Section~\ref{sec:method} we outline the modified methodology of interleaved data layout for RHS matrics and a single global copy for the LHS matrix.
In Sections~\ref{sec:tri} and~\ref{sec:pent} we outline the specifics for tridiagonal and pentadiagonal matrices respectively including also benchmarks against the state of the art algorithms to show performance is at least as good, if not better in most cases of batch size and matrix size. 
Finally we present our conclusions in Section~\ref{sec:con}.


\begin{figure}
\begin{tikzpicture}[
    grow=right,
    level 1/.style={sibling distance=5.0cm,level distance=5.0cm},
    level 2/.style={sibling distance=2.5cm, level distance=4cm},
    level 3/.style={sibling distance=1.0cm, level distance=4cm},
    edge from parent/.style={very thick,draw=blue!40!black!60},
    edge from parent path={(\tikzparentnode.east) -- (\tikzchildnode.west)},
    kant/.style={text width=2cm, text centered, sloped},
    every node/.style={text ragged, inner sep=2mm},
    punkt/.style={rectangle, rounded corners, shade, top color=white,
    bottom color=blue!50!black!20, draw=blue!40!black!60, very
    thick }
    ]

\node[punkt, text width=5.5em] {\textbf{Device RAM}}
    child 
    {
        node [punkt]{\textbf{L2 Cache} \nodepart{second}}
        child 
	    {
	        node [punkt]{\textbf{L1 Cache} \nodepart{second}}
	        child 
		    {
		        node [punkt]{\textbf{Warp} \nodepart{second}}
		        edge from parent
		        node[kant, below, pos=.6] {}
		    }
		    child 
		    {
		        node [punkt]{\textbf{Warp} \nodepart{second}}
		        edge from parent
		        node[kant, below, pos=.6] {}
		    }
		    child 
		    {
		        node [punkt]{\textbf{Warp} \nodepart{second}}
		        edge from parent
		        node[kant, below, pos=.6] {}
		    }
		    child 
		    {
		        node [punkt]{\textbf{Warp} \nodepart{second}}
		        edge from parent
		        node[kant, below, pos=.6] {}
		    }
	        edge from parent
	        node[kant, below, pos=.6] {}
	    }
        child 
	    {
	        node [punkt]{\textbf{L1 Cache} \nodepart{second}}
	        edge from parent
	        node[kant, below, pos=.6] {}
	    }
        edge from parent
        node[kant, below, pos=.6] {}
    }
    child 
    {
        node [punkt]{\textbf{L2 Cache} \nodepart{second}}
        child 
	    {
	        node [punkt]{\textbf{L1 Cache} \nodepart{second}}
	        edge from parent
	        node[kant, below, pos=.6] {}
	    }
        child 
	    {
	        node [punkt]{\textbf{L1 Cache} \nodepart{second}}
	        child 
		    {
		        node [punkt]{\textbf{Warp} \nodepart{second}}
		        edge from parent
		        node[kant, below, pos=.6] {}
		    }
		    child 
		    {
		        node [punkt]{\textbf{Warp} \nodepart{second}}
		        edge from parent
		        node[kant, below, pos=.6] {}
		    }
		    child 
		    {
		        node [punkt]{\textbf{Warp} \nodepart{second}}
		        edge from parent
		        node[kant, below, pos=.6] {}
		    }
		    child 
		    {
		        node [punkt]{\textbf{Warp} \nodepart{second}}
		        edge from parent
		        node[kant, below, pos=.6] {}
		    }
	        edge from parent
	        node[kant, below, pos=.6] {}
	    }    
        edge from parent
        node[kant, below, pos=.6] {}
    };
\end{tikzpicture}
\caption{Diagram of memory hierarchy in an NVIDIA GPU. We omit nodes for the purposes of the diagram, the top and bottom branches are representative of the memory routing from Device Ram to warp. When every thread is accessing the same memory location warps which share the same SM will receive an L1 cache hit, SMs sharing an L2 cache will receive an L2 cache hit and if the data is not present in the L2 cache it will be retrieved from the Device RAM. Thus it can be seen here how the first warp to request a location in memory in a given group will be the slowest but accesses for following warps become faster and faster as the memory propagates down the tree into various caches.}
\label{fig:memoryaccess}
\end{figure}
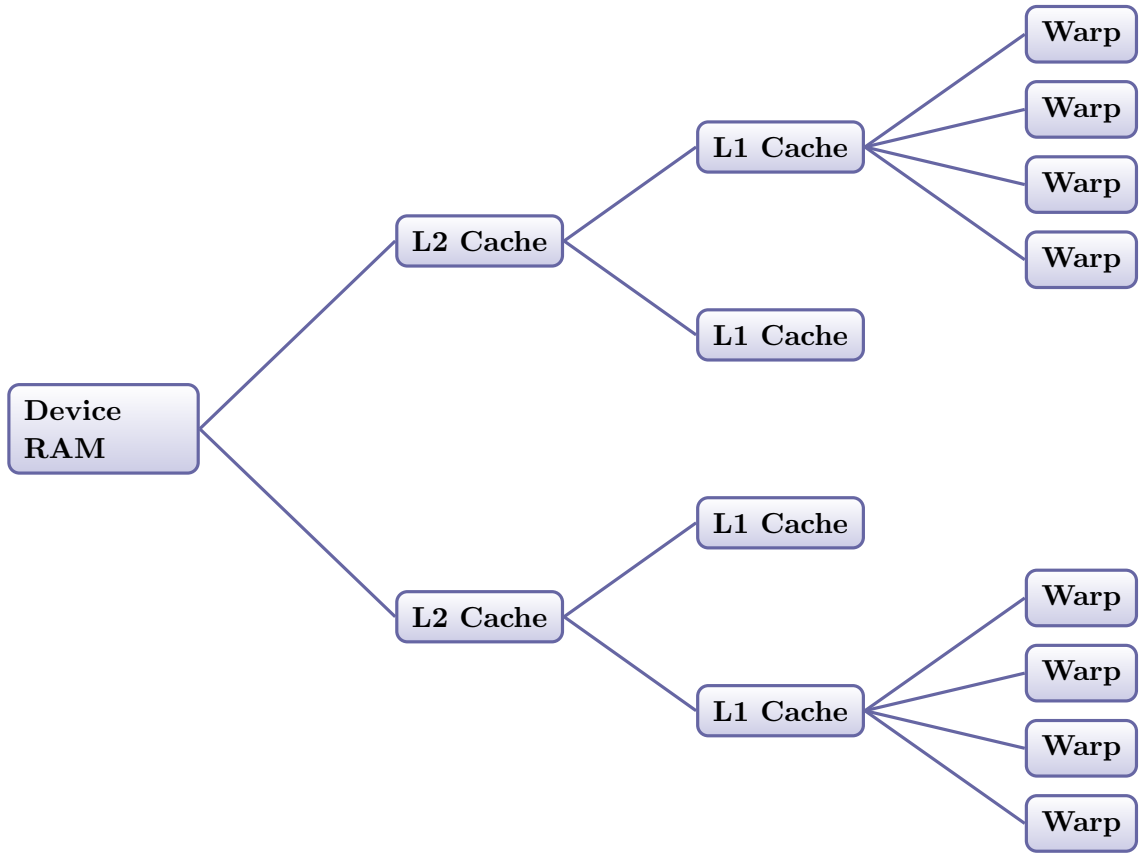

\section{Matrix Solver Methodology}
\label{sec:method}
The methodologies implementing the tridiagonal and pentadiagonal algorithms we are concerned with in this paper have to date relied on using interleaved data layouts in global memory.
Each thread handles a particular system solution and the system is solved by a forward sweep followed by a backward sweep.  
The interleaved data format is to ensure coalescence of data accesses during theses sweeps, with each thread retrieving neighbouring memory locations in global memory for various LHS and RHS matrix entries~\cite{cuThomasBatch}.
This maximises bandwidth but is limiting to the overall memory budget as each system requires its own copy of the LHS matrix.
Instead we propose to store globally just a single copy of the LHS matrix entries with all threads accessing the desired value at the same time, the RHS matrices will still be stored in the same interleaved data access pattern as before.
This will reduce the overall bandwidth of memory accesses to the LHS but this will not harm the solver's performance as we will show in later sections.
Critically also this approach of using just a single LHS matrix will save drastically on the amount of memory used in batch solvers allowing for larger batch numbers and greater matrix sizes to be solved on a single GPU, increasing hardware usage efficiency.

We now discuss the memory access pattern for the single LHS matrix.
Global memory access in GPUs is achieved via the L2 cache.
When a piece of memory is requested for by a warp this memory is retrieved from RAM and copied to the L2 cache followed by the L1 cache on the SM before then being used by the relevant warps being computed on that SM.
Each SM has their own pipe to the L2 cache. 
So when warps from different SMs all request the same memory they will all get a cache hit in at least the L2 cache except the first one to arrive which will be the warp to retrieve the data from the RAM on the device.
We can guarantee that this will occur as each warp will be computing the same instruction in either the forward or backward sweep of the solver but warps on different SMs will not arrive to the L2 cache at the same time.
In addition to this warps which share L1 caches will get cache hits here when they're not the first to request the piece of memory. 
A diagram of this memory access pattern discussion can also be seen in Figure~\ref{fig:memoryaccess}.
Thus it is clear that given every piece of memory must follow the above path most of the threads solving a batched problem will benefit from speed--ups due to cache hits as they no longer need to retrieve their own copy of the data.

In the following two sections we present specific details for each of the tridiagonal and pentadiagonal schemes along with the previously discussed benchmarks.
The benchmarks are performed on 16GB Tesla V100 GPUs with the pre-factorisation step for the single LHS performed on the CPU.
We shall refer to the new versions of the algorithms as cuThomasConstantBatch and cuPentConstantBatch for tridiagonal and pentadiagonal respectively.
We use the nomenclature 'Constant' denoting the fact that all systems have the same LHS. 
The RHS will be stored as usual in an interleaved data format and is computed using cuSten, an in--house finite--differnce library for GPUs~\cite{gloster2019custen}.
For notational purposes we will use $N$ to describe the number of unknowns in our systems and $M$ to describe the batch size (number of systems being solved).


\section{Tridiagonal Systems}
\label{sec:tri}
In this section we first present the Thomas Algorithm~\cite{numalgC} and how it is modified for batch solutions with multiple RHS matrices and with a single LHS. 
We then present our chosen benchmark problem of periodic diffusion equations and then the results.

\subsection{Tridiagonal Inversion Algorithm}
We begin with a generalised tridiagonal matrix system $\mxA \vecx=\vecrhsT$ with $\mxA$ given by
\begin{equation*}
\mxA = 
\begin{pmatrix}
b_1 & c_1 & 0 & \cdots &   \cdots & 0 \\
a_2 & b_2 & c_2 & 0 & \cdots   & \vdots \\
0 & a_3 & b_3 & c_3 & 0 &  \vdots \\
\vdots & \ddots & \ddots & \ddots & \ddots & 0  \\
\vdots & \cdots  & 0   & a_{N - 1}  & b_{N - 1} & c_{N - 1} \\
0 & \cdots  &   \cdots & 0 & a_{N}  & b_{N} 
\end{pmatrix}.
\end{equation*}
We then solve this system using a pre-factorisation step followed by a forwards and backwards sweep. 
The pre-factorisation is given by
\begin{equation}
\hat{c}_1 = \frac{c_1}{b_1}
\end{equation}
And then for $i = 2, \dots, N$
\begin{equation}
\hat{c}_i = \frac{c_i}{b_i - a_i \hat{c}_{i - 1}}
\end{equation}
For the forwards sweep we have
\begin{equation}
\hat{d}_1 = \frac{d_1}{b_1}
\end{equation}
\begin{equation}
\hat{d}_i = \frac{d_i - a_i \hat{d}_{i - 1}}{b_i - a_i \hat{c}_{i - 1}}
\end{equation}
While for the backwards sweep we have 
\begin{equation}
x_N = \hat{d}_N
\end{equation}
And for $i = N-1, \dots, 1$
\begin{equation}
x_i = \hat{d}_i - a_i \hat{c}_i
\end{equation}

Previous applications of this algorithm~\cite{cuThomasBatch} required that each thread had access to its own copy of $4$ vectors, the $3$ diagonals $a_i$, $b_i$ and $c_i$ along with the the RHS $d_i$.
These would then be overwritten in the pre-factorisation and solve steps to save memory, thus the total memory usage here is $O(4 \times M \times N)$.
We now limit the LHS to a single global case that will be accessed simultaneously using all threads as discussed in Section~\ref{sec:method} and retain the individual RHS in interleaved format for each thread $f_i$.
This reduces the data storage to $O(3 \times N + M * N)$, an approximate 75\% reduction.
We present benchmark methodology and results for this method in the following subsections.

\subsection{Benchmark Problem}
For a benchmark problem we solve the diffusion equation, a standard model equation in Computational Science and Engineering, its presence can be seen in most systems involving heat and mass transfer.
The solution as $t \rightarrow \infty$ is also a solution of a Poisson equation.
The equation in one dimension is given as 
\begin{equation}
\frac{\partial C}{\partial t} = \alpha \frac{\partial^2 C}{\partial x^2} 
\label{eq:diffusion}
\end{equation} 
where $\alpha$ is the diffusion coefficient.
We solve this equation on a periodic domain of length $L$ such that $C(x + L) = C(x)$ with an initial condition $C(x, t = 0) = f(x)$ valid on the domain.
We rescale by setting $\alpha = 1$ and $L = 1$ and integrate Eq.~\eqref{eq:diffusion} in time using a standard Crank--Nicholson scheme with central differences for space which is unconditionally stable.
Finite differencing is done using standard notation
\begin{equation}
C_i^n = C(x=i\Delta x ,t=n\Delta t)
\label{eq:stdDiff}
\end{equation}
where $\Delta x = L / N$ and $i = 1 \dots N$.
Thus our numerical scheme can be written as 
\begin{equation}
- \sigma_x C_{i - 1}^{n+1} + (1 + 2 \sigma_x) C_{i}^{n+1} - \sigma_x C_{i+1}^{n+1}
=  \sigma_x C_{i - 1}^{n} + (1 - 2 \sigma_x)C_{i}^{n} + \sigma_x C_{i+1}^{n} 
\label{eq:1ddiffscheme}
\end{equation} 
where
\begin{equation}
\sigma_x = \frac{\Delta t}{2 \Delta x^2}
\end{equation}
Thus we can relate these coefficients to matrix entries by
\begin{equation}
a_i  = - \sigma_x,\qquad b_i = 1 + 2 \sigma_x,\qquad
c_i = - \sigma_x
\end{equation}%

In the following section we present our method to deal with the periodicity of the matrix and then after that present the benchmark comparison with the existing state of the art.

\subsection{Periodic Tridiagonal Matrix}
As the system is periodic two extra entries will appear in the matrix, one in the top right corner and another in the bottom left, thus our matrix is now given by
\begin{equation*}
\mxA = 
\begin{pmatrix}
b & c & 0 & \cdots &  \cdots & 0 & a \\
a & b & c & 0 & \cdots & \cdots   & 0 \\
0 & a & b & c & 0 & \cdots  & \vdots \\
\vdots & \ddots & \ddots & \ddots & \ddots & \ddots &   \vdots\\
\vdots & \cdots  & 0  &  a  & b  & c & 0  \\
0 & \cdots & \cdots  & 0   & a  & b & c \\
c & 0 & \cdots  &   \cdots & 0 & a  & b 
\end{pmatrix}.
\end{equation*}
In order to deal with these we use the Sherman-Morrison formula.
We rewrite our system as 
\begin{equation}
\mxA\vecx  = (\mxA' + \mxU \otimes \mxV) \vecx = \vecrhsT 
\end{equation} 
Where 
\begin{subequations}
\begin{equation}
\mxU = 
\begin{pmatrix}
-b \\
0 \\
\vdots \\
\vdots \\
0 \\
c\\
\end{pmatrix},
\qquad
\mxV = 
\begin{pmatrix}
1 \\
0 \\
\vdots \\
\vdots \\
0 \\
- a / b\\
\end{pmatrix},
\qquad
\end{equation}
\begin{equation}
\mxA' = 
\begin{pmatrix}
2b & c & 0 & \cdots &  \cdots & 0 & a \\
a & b & c & 0 & \cdots & \cdots   & 0 \\
0 & a & b & c & 0 & \cdots  & \vdots \\
\vdots & \ddots & \ddots & \ddots & \ddots & \ddots &   \vdots\\
\vdots & \cdots  & 0  &  a  & b  & c & 0  \\
0 & \cdots & \cdots  & 0   & a  & b & c \\
c & 0 & \cdots  &   \cdots & 0 & a  & b + a c / b
\end{pmatrix}
\end{equation}%
\end{subequations}%
Thus two tridiagonal systems must now be solved
\begin{equation}
\mxA' \vecy = \vecrhsT  \qquad \mxA' \vecx = \mxU
\end{equation}
the second of which need only be performed once at the beginning of a given simulation.
Finally to recover $\vecx$ we substitute these results into 

\begin{equation}
\vecx = \vecy - \left(\frac{\mxV \cdot \vecy}{1 + (\mxV \cdot \vecz)}\right)\vecz
\end{equation}

\begin{figure}[h]
	\centering
		\includegraphics[width=0.8\textwidth]{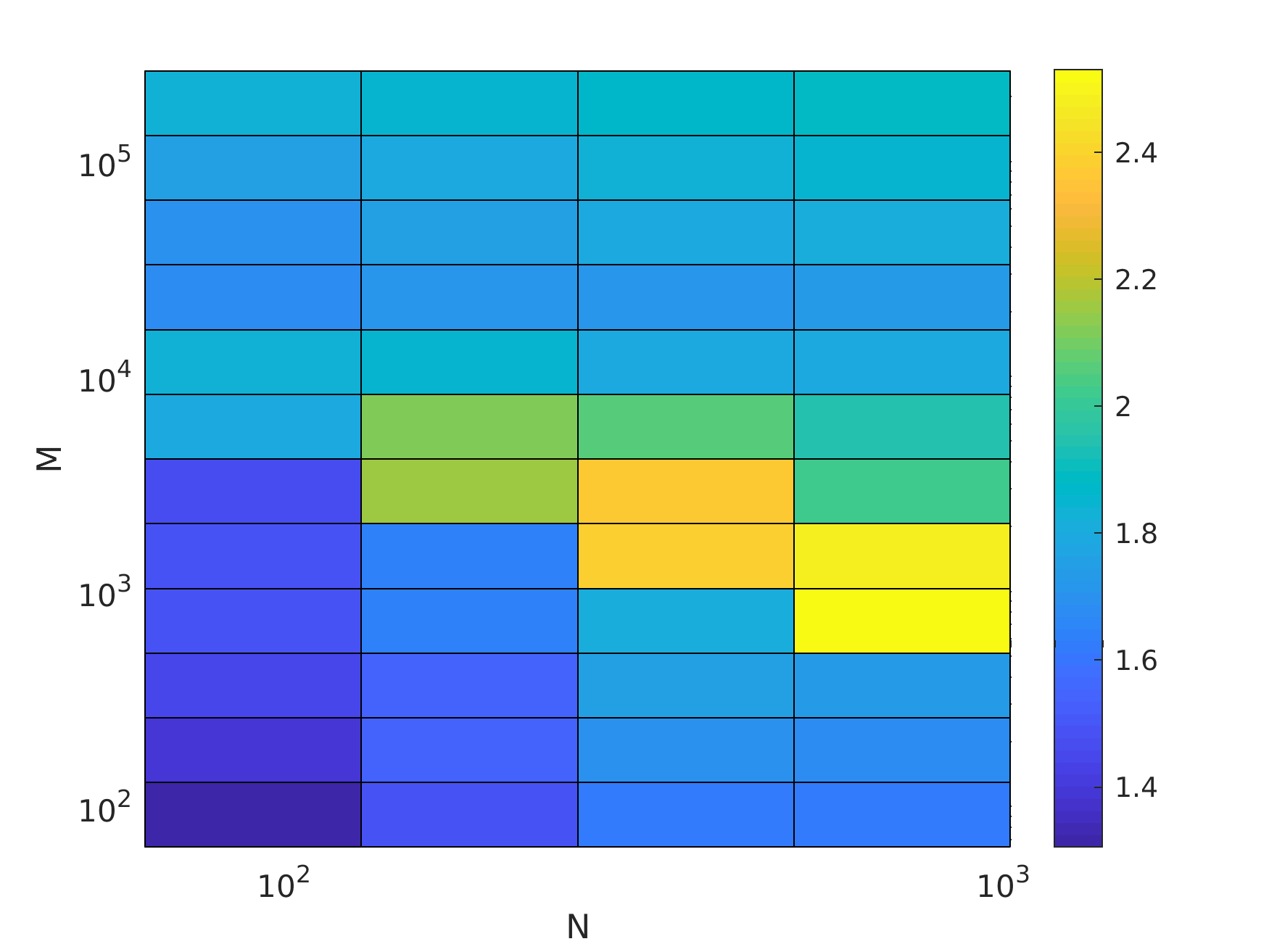}
		\caption{Speedup of cuThomasConstantBatch versus cuThomasBatch (gtsvInterleavedBatch).}
	\label{fig:speedupInterleaved}
\end{figure}

\subsection{Benchmark Results}
The diffusion equation benchmark problem presented above is solved for 1000 time--steps in order to extract the relevant timing statistics and average out any effects of background processes on the benchmark.
Both codes are timed for just the time--stepping sections of the process, we omit start up costs such as memory allocation, setting of parameters etc.
The speed--up results can be seen in Figure~\ref{fig:speedupInterleaved}, in all situations there is a clear speed--up over the existing state of the art cuThomasBatch. 
The largest speed--ups are for $N$ large and moderate $M$.
Significant speed--up is available for all of large $M$ which is the domain where GPUs would typically be deployed to solve the problem.

It should be noted that some of the speed--up seen here can be attributed to the pre--factorisation step completed by the authors which the cuThomasBatch implementation does not have. 
cuThomasBatch requires that the LHS matrix be reset at every time--step as the pre--factorisation and solve steps are carried out in the one function, leading to an overwrite of the data and making it unusable for repeated time--stepping.
This overwriting feature has been seen in previous studies~\cite{cuPentBatch} where the authors also carried out a rewrite to make the benchmarking conditions between their work and the state of the art.
It was shown that not all of the speed--up can be attributed to the lack of needing to reset the LHS matrix, thus some of the performance increase we are seeing in Figure~\ref{fig:speedupInterleaved} can be attributed to the new data layout presented in this paper.
Our second benchmark for the pentadiagonal case will also show this as there was no resetting of the matrix required.


\section{Pentadiagonal Systems}
\label{sec:pent}
In this section we first present a modified version of the pentadiagonal inversion algorithm as presented in~\cite{cuPentBatch} with a single LHS. 
Then we present the benchmark problem and finally this is followed by the results comparing our new implementation with existing state of the art.

\subsection{Pentadiagonal Inversion Algorithm}
In this section we describe a standard numerical method~\cite{numalgC, cuPentBatch} for solving a pentadiagonal problem $\mxA\vecx=\vecrhs$.  We present the algorithm in a general context so we have a pentadiagonal matrix given by

\begin{equation*}
\mxA = 
\begin{pmatrix}
c_1 & d_1 & e_1 & 0 & \cdots &  0 & \cdots & 0 \\
b_2 & c_2 & d_2 & e_2 & 0 & \cdots & \cdots   & \vdots \\
a_3 & b_3 & c_3 & d_3 & e_3 & 0 & \cdots  & 0\\
0 & \ddots & \ddots & \ddots & \ddots & \ddots & \ddots &   \vdots\\
\vdots& \ddots & \ddots & \ddots & \ddots & \ddots & \ddots & 0 \\
0 & \cdots  & 0  &  a_{N - 2}  & b_{N - 2}  & c_{N - 2} & d_{N - 2} & e_{N - 2} \\
0 & \cdots & \cdots  & 0   & a_{N - 1}  & b_{N - 1} & c_{N - 1} & d _{N - 1}\\
0 & \cdots & \cdots  &   \cdots & 0 &  a_{N} & b_{N}  & c_{N} 
\end{pmatrix}.
\end{equation*}
Three steps are required to solve the system:
\begin{enumerate}
\item Factor $\mxA = \mxL\mxR$  to obtain $\mxL$ and $\mxR$.
\item Find $\vecg$ from $\vecf = \mxL\vecg$
\item Back-substitute to find $\vecx$ from $\mxR\vecx = \vecg$
\end{enumerate}
Here, $\mxL$, $\mxR$ and $\vecg$ are given by the following equations:
\begin{subequations}
\begin{equation}
\mxL = 
\begin{pmatrix}
\alpha_1 &  &  &  &  &   &   \\
\beta_2 & \alpha_2 &  &  &  &  &     \\
\epsilon_3 & \beta_3 & \alpha_3 &  &  &  &   \\
 & \ddots & \ddots & \ddots &  &  &     \\
 &  &  \epsilon_{N - 1}  & \beta_{N - 1}  & \alpha_{N - 2} &    \\
&  &    & \epsilon_{N - 1}  & \beta_{N - 1} & \alpha_{N }\\
\end{pmatrix},
\qquad
\vecg = \begin{pmatrix}
g_{1} \\
g_{2} \\
\vdots \\
\vdots \\
g_{N-1} \\
g_{N}
\end{pmatrix},
\end{equation}
\begin{equation}
\mxR = 
\begin{pmatrix}
1 & \gamma_1  & \delta_1  &  &  &   &   \\
 & 1 & \gamma_2  & \delta_2  &  &  &     \\
 &  & \ddots & \ddots & \ddots  &  &   \\
 & & & 1  & \gamma_{N-2}  & \delta_{N-2}    \\
 &  &  &  & 1 & \gamma_{N-1}    \\
&  &    & &  & 1\\
\end{pmatrix}
\end{equation}%
\end{subequations}%
(the other entries in $\mxL$ and $\mxR$ are zero).  
The explicit factorisation steps for the factorisation $\mxA=\mxL\mxR$ are as follows:
\begin{enumerate}
\item $\alpha_1 = c_1$
\item $\gamma_1 = \frac{d_1}{\alpha_1}$
\item $\delta_1 = \frac{e_1}{\alpha_1}$
\item $\beta_2 = b_2$
\item $\alpha_2 = c_2 - \beta_2\gamma_1$
\item $\gamma_2 = \frac{d_2 - \beta_2 \delta_1}{\alpha_2}$
\item $\delta_2 = \frac{e_2}{\alpha_2}$
\item For each $i = 3, \dots, N-2$
\begin{enumerate}[label*=\arabic*.]
\item $\beta_i = b_i - a_i \gamma_{i-2}$
\item $\alpha_i = c_i - a_i\delta_{i-2} - \beta_i \gamma_{i-1}$
\item $\gamma_i = \frac{d_i - \beta_i \delta_{i-1}}{\alpha_i}$
\item $\delta_i = \frac{e_i}{\alpha_i}$
\end{enumerate}
\item $\beta_{N-1} = b_{N-1} - a_{N - 1}\gamma_{N-3}$
\item $\alpha_{N - 1} =  c_{N-1} - a_{N-1}\delta_{N-3} - \beta_{N-1}\gamma_{N-2}$
\item $\gamma_{N-1} = \frac{d_{N-1}-\beta_{N-1}\delta_{N-2}}{\alpha_{N-1}}$
\item $\beta_{N} = b_{N} - a_{N }\gamma_{N-2}$
\item $\alpha_{N} =  c_{N}- a_{N}\delta_{N-2} - \beta_{N}\gamma_{N-1}$
\item $\epsilon_i = a_i, \quad \forall i$
\end{enumerate}
The steps to find $\vecg$ are as follows:
\begin{enumerate}
\item $g_1 = \frac{f_1}{ \alpha_1}$
\item $g_2 = \frac{f_2 - \beta_2 g_1}{\alpha_2}$
\item  $g_i = \frac{f_i - \epsilon_i g_{i-2} - \beta_i g_{i - 1}}{\alpha_i} \quad \forall i = 3 \cdots N$
\end{enumerate}
Finally, the back-substitution steps  find $\vecx$ are as follows:
\begin{enumerate}
\item $x_N = g_N$
\item $x_{N-1} = g_{N-1} - \gamma_{N-1}x_N$
\item  $x_i = g_i - \gamma_i x_{i+1} - \delta_{i}x_{i+2} \quad \forall i = (N-2) \cdots 1$
\end{enumerate}

Previous applications of this algorithm~\cite{cuPentBatch} required that each thread had access to its own copy of $6$ vectors, the $5$ diagonals $a_i$, $b_i$, $c_i$, $d_i$ and $e_i$ along with the RHS $f_i$.
These would then be overwritten in the pre-factorisation and solve steps to save memory, thus the total memory usage here is $O(6 \times M \times N)$.
We now limit the LHS to a single global case that will be accessed simultaneously using all threads as discussed in Section~\ref{sec:method} and retain the individual RHS in interleaved format for each thread $f_i$.
This reduces the data storage to $O(5 \times N + M \times N)$, this is an approximate 83\% reduction in data usage.
We present benchmark methodology and results for this method in the following subsections.
For completion we present results for an extra implementation where all the entries on respective diagonals are equal eliminating the need to store $a_i$ (which is also $\epsilon_i$), reducing the storage further to $O(4 \times N + M \times N)$, we shall refer to these results as cuPentUniformBatch.

\subsection{Benchmark Problem}
As in the paper \cite{cuPentBatch} we will solve batches of the periodic hyperdiffusion equation in 1D, a key equation in solving a Cahn--Hilliard like system, as a benchmark problem.
The method is also representative of a 4th order accurate Laplace system.
The hyperdiffusion equation in 1D is given as 
\begin{equation}
\frac{\partial C}{\partial t}=- D \frac{\partial^4 C}{\partial x^4},\qquad t>0,\qquad x\in (0,L),
\label{eq:hyperdiff}
\end{equation}
with periodic boundary condition $C(x+L,t)=C(x)$ and initial condition $C(x,t=0)=f(x)$, valid on $[0,L]$.  
We rescale by setting $L=D=1$.  We discretize Equation~\eqref{eq:hyperdiff} in space using centred differences and in time using the Crank--Nicholson method which is unconditionally stable.  
We use same standard notation for the discretisation as in Eq.~\eqref{eq:stdDiff}   
In this way, the discretised version of Equation~\eqref{eq:hyperdiff} is written as
\begin{multline}
\frac{C_i^{n+1}-C_i}{\Delta t}=
-\tfrac{1}{2}\Delta x^{-4}\left[C_{i+2}^{n+1}-4C_{i+1}^{n+1}+6C_{i}^{n+1}-4C_{i-1}^{n+1}+C_{i-2}^{n+1}\right]\\
-\tfrac{1}{2}\Delta x^{-4}\left[C_{i+2}^{n}  -4C_{i+1}^{n}  +6C_{i}^{n}  -4C_{i-1}^{n}+C_{i-2}^{n}\right].
\label{eq:disc}
\end{multline}
Upon rearranging terms, Equation~\eqref{eq:disc} can be written more compactly as follows:
\begin{multline}
\sigma_x C_{i - 2}^{n + 1} - 4 \sigma_x C_{i - 1}^{n+1} + (1 + 6 \sigma_x)C_{i}^{n+1} - 4 \sigma_x C_{i+1}^{n+1} + \sigma_x C_{i+2}^{n+1} \\ 
=  - \sigma_x C_{i - 2}^{n} + 4 \sigma_x C_{i - 1}^{n} + (1 - 6 \sigma_x)C_{i}^{n} + 4 \sigma_x C_{i+1}^{n} - \sigma_x C_{i+2}^{n},
\label{eq:1d_hyper_scheme}
\end{multline} 
where $\sigma_x = \Delta t / 2\Delta x^4$. 
Thus we can clearly relate these finite difference equations back to our matrix system by
\begin{subequations}
\begin{equation}
a_i  = \sigma_x,\qquad b_i = - 4 \sigma_x,\qquad
c_i = 1 + 6\sigma_x, \qquad
d_i = - 4 \sigma_x, \qquad e_i = \sigma_x.
\end{equation}%
Similarly,
\begin{equation}
f_i = - \sigma_x C_{i - 2}^{n} + 4 \sigma_x C_{i - 1}^{n} + (1 - 6 \sigma_x)C_{i}^{n} + 4 \sigma_x C_{i+1}^{n} - \sigma_x C_{i+2}^{n}.
\end{equation}%
\label{eq:matrix_sys}%
\end{subequations}%
The method for dealing with the periodic boundaries is the same as that presented in~\cite{cuPentBatch}.
\begin{figure}[h]
	\centering
		\includegraphics[width=0.8\textwidth]{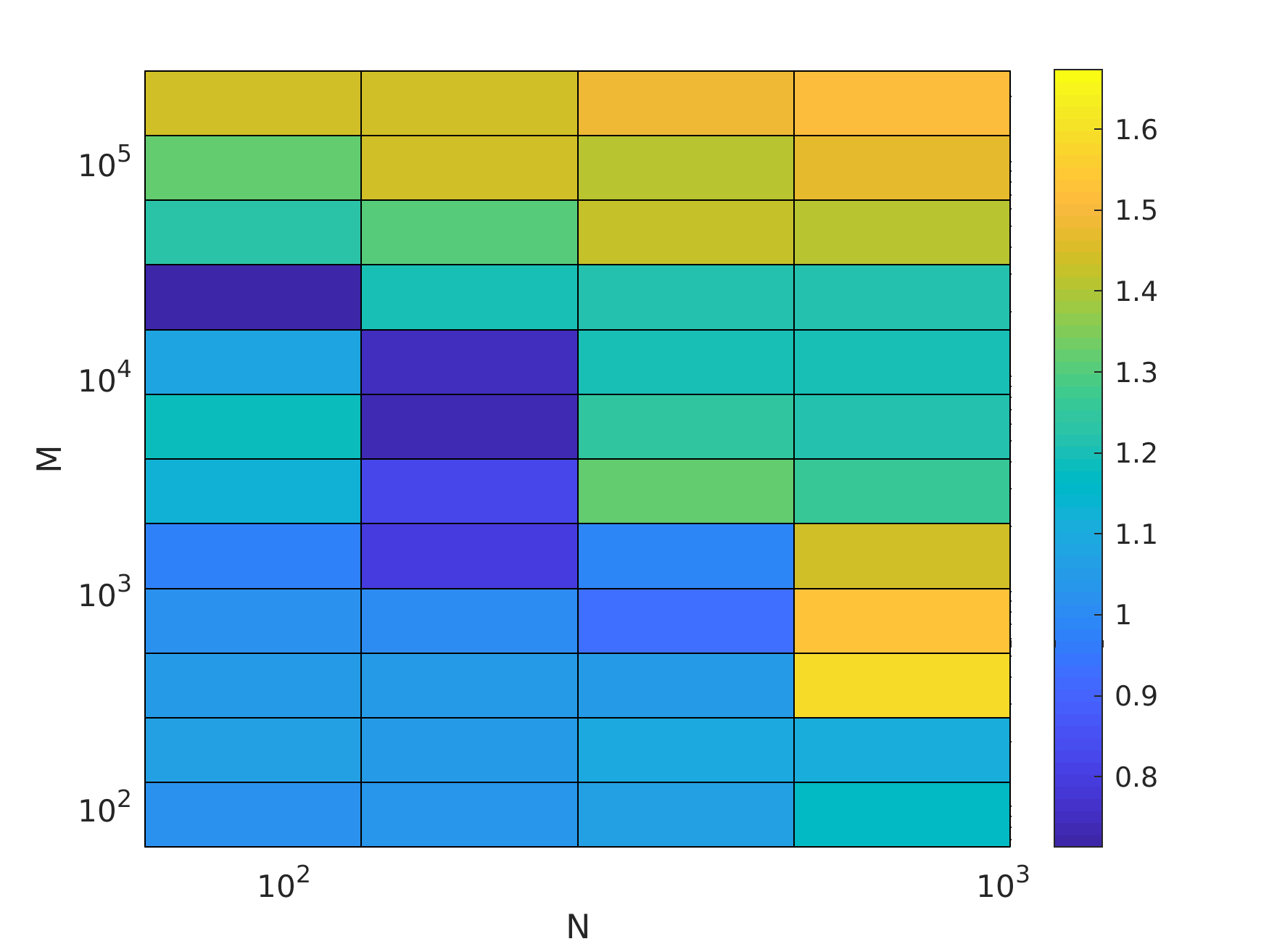}
		\caption{Speedup of cuPentConstantBatch versus cuPentBatch.}
	\label{fig:pentConstantSpeedUp}
\end{figure}
\subsection{Benchmark Results}
We plot the speed--up of cuPentConstantBatch versus cuPentBatch solving batches of the above method for the hyperdiffusion equation in Figure~\ref{fig:pentConstantSpeedUp}. 
In order to calculate the values for $\vecrhs$ we use cuSten \cite{gloster2019custen}.
It can be seen in the figure that cuPentBatchConstant outperforms cuPentBatch consistently for high values of both $M$ and $N$. 
At low $M$ and $N$ they are roughly equivalent and there are some areas where cuPentBatch performs better, generally a standard CPU implementation is more desirable at these low numbers when the movement over the PCI lane is taken into account along with other overheads so we can discount these. 
If we push the values out further than those plotted the difference becomes orders of magnitudes as the memory for cuPentBatch rapidly exceeds the available RAM on the GPU while cuPentBatchConstant can still fit.

\begin{figure}[h]
	\centering
		\includegraphics[width=0.8\textwidth]{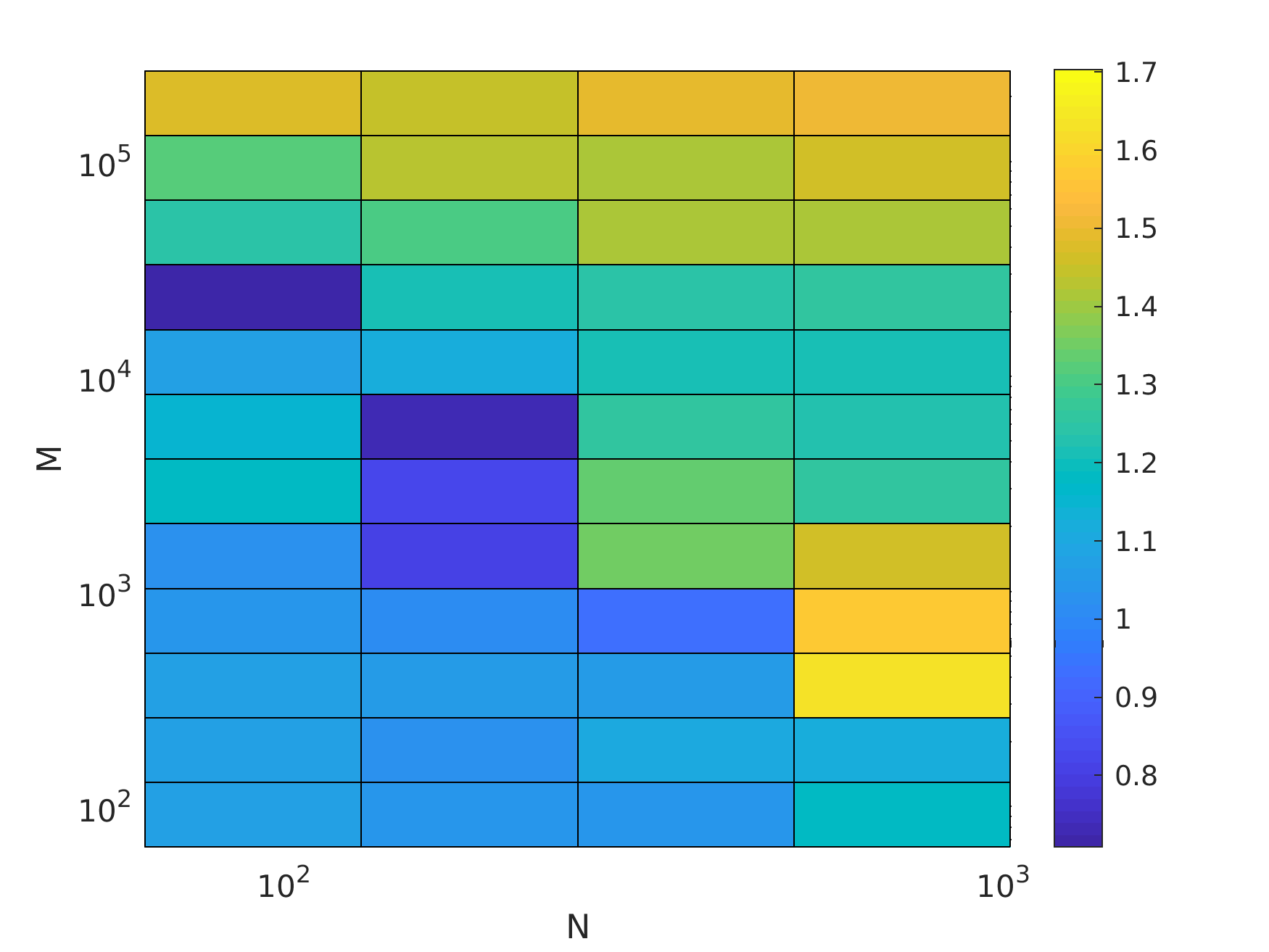}
		\caption{Speed--up of cuPentUnifromBatch versus cuPentBatch.}
	\label{fig:pentUniformSpeedUp}
\end{figure}

Thus we conclude in situations where there is uniform LHS matrices with multiple RHS and the algorithm is suitably stable for the problem (symmetric positive definite is enough here) that cuPentBatchConstant is a better choice in terms of both memory usage (allowing for larger values of $M$ and $N$ on one GPU) and speed.
Similar results can be seen in Figure~\ref{fig:pentUniformSpeedUp} for the case of cuPentUniformBatch with slight improvements in overall speed in certain locations due to the lack of access to the vector storing $\epsilon_i$.
Slight advantages of cuPentUniformBatch over cuPentConstantBatch are apparent where it can be used, if the functions were being repeatedly called enough times in a given simulation its use is certainly warranted as the savings on time will compound with each call. 

\section{Conclusions}
\label{sec:con}
We have shown that an interleaved batch of RHS matrices along with a single LHS achieves large reductions in data usage along with a substantial speed--up when compared with the existing state of the art.
The reduction in data usage allows for greater use of hardware resources to be made with significant extra space that can now be devoted to solving more systems of equations simultaneously rather than unnecessarily storing data.
The extra equations now being solved by one GPU along with the speed--ups provided by the new implementation are a significant improvement over the state of the art in applications where only one LHS matrix is required for the solution of all the systems in the batch.
In addition the reduction in the number of GPUs one would need when solving very large batches, coupled with the speed--ups achieved, leads to a reduction in electricity usage.
As HPC moves into the future, energy needs are becoming more and more prevalent, the increases in FLOPs/WATT provided by GPUs have both financial and environmental implications.
Thus there is an increasing need for the use of GPUs in HPC platforms and the need for efficient algorithms such as those presented in this paper for use on them.


\section*{Acknowledgements}
Andrew Gloster acknowledges funding received from the UCD Research Demonstratorship.   
Enda Caroll acknowledges funding recieved under the Government of Ireland Postgraduate Scholarship Programme funded by Irish Research Council, grant number GOIPG/2018/2653.
All authors gratefully acknowledge the support of NVIDIA Corporation with the donation of the Titan X Pascal GPUs used for this research.  
The authors also thank Lung Sheng Chien for helpful discussions throughout this project.
The authors wish to acknowledge the DJEI/DES/SFI/HEA Irish Centre for High-End Computing (ICHEC) for the provision of computational facilities and support.

\section*{References}

\bibliographystyle{unsrt}
\bibliography{cpcbiblio}

\end{document}